\documentclass[prb,aps,preprint,amsmath,amssymb,showpacs,superscriptaddress,floatfix]{revtex4-1}

\usepackage[colorlinks,citecolor=blue,linkcolor=blue]{hyperref}
\usepackage{graphicx}
\usepackage{dcolumn}
\usepackage{bm}
\usepackage{subcaption}
\UseRawInputEncoding

\DeclareMathOperator{\sinc}{sinc}

\begin{document}

\title{Asymmetric spin wave dispersion due to a saturation magnetization gradient}

\author{P. Borys}
\email{pabloborys@ciencias.unam.mx}
\affiliation{Instituto de Ciencias Aplicadas y Tecnolog\'ia, Universidad Nacional Aut\'onoma de M\'exico, Ciudad Universitaria 04510, Mexico}
\affiliation{Department of Physics and Atmospheric Science, Dalhousie University, Halifax, Nova Scotia, Canada B3H 4R2}
\author{O. Kolokoltsev}
\affiliation{Instituto de Ciencias Aplicadas y Tecnolog\'ia, Universidad Nacional Aut\'onoma de M\'exico, Ciudad Universitaria 04510, Mexico}
\author{N. Qureshi}
\affiliation{Instituto de Ciencias Aplicadas y Tecnolog\'ia, Universidad Nacional Aut\'onoma de M\'exico, Ciudad Universitaria 04510, Mexico}
\author{M. L. Plumer}
\affiliation{Department of Physics and Atmospheric Science, Dalhousie University, Halifax, Nova Scotia, Canada B3H 4R2}
\affiliation{Department of Physics and Physical Oceanography, Memorial University of Newfoundland, St. John's, Newfoundland and Labrador, Canada A1B 3X7}
\author{T. L. Monchesky}
\affiliation{Department of Physics and Atmospheric Science, Dalhousie University, Halifax, Nova Scotia, Canada B3H 4R2}

\date{\today}

\begin{abstract}
	We demonstrate using micromagnetic simulations and a theoretical model that a gradient in the saturation magnetization ($M_s$) of a perpendicularly magnetized ferromagnetic film induces a non-reciprocal spin wave propagation and, consequently an asymmetric dispersion relation. The $M_s$ gradient adds a linear potential to the spin wave equation of motion consistent with the presence of a force. We consider a transformation from an inertial reference frame in which the $M_s$ is constant to an accelerated reference frame where the resulting inertial force corresponds to the force from the $M_s$ gradient. As in the Doppler effect, the frequency shift leads to an asymmetric dispersion relation. Additionally, we show that under certain circumstances, unidirectional propagation of spin waves can be achieved which is essential for the design of magnonic circuits. Our results become more relevant in light of recent experimental works in which a suitable thermal landscape is used to dynamically modulate the saturation magnetization. 	
\end{abstract}
	
	
\maketitle
	
	%
	
	Spin waves, collective excitations in magnetic media, transport information without any particle motion, e.g., electric charge, and are hence free of the undesired Joule heating. Magnonics -the field that study the behavior of spin waves and their quanta magnons, has received much attention as a plausible complement to conventional semiconductor electronics for data transport and processing\cite{chen_strong_2018,chumak_magnon_2015,csaba_perspectives_2017,grundler_reconfigurable_2015,khitun_magnonic_2010,kruglyak_magnonics_2010,neusser_magnonics:_2009,schneider_realization_2008,yu_magnon_2018}.  For magnonic devices to be relevant, they need to be miniaturized to the nanoscale which in turn needs spin wave wavelengths on the nanometer scale where the exchange interaction dominates over dipole/magnetostatic energies. It was only recently that excitation and measurement of exchange spin waves in films was finally possible opening a wide range of technological paths\cite{che_efficient_2020,hamalainen_tunable_2017,Heinz2020,liu_long-distance_2018}. Contrary to long wavelength, magnetostatic dominated spin waves that can exhibit non-reciprocal propagation, exchange spin waves are isotropic in their propagation due to the quadratic form of its dispersion relation. Several mechanisms have been proposed to make the dispersion asymmetric since anisotropic propagation is key to the design of magnonic circuitry. Examples of such mechanisms include, induced Dzyaloshinskii-Moriya Interaction\cite{garcia-sanchez_nonreciprocal_2014,belmeguenai_interfacial_2015,di_asymmetric_2015,moon_spin-wave_2013,dosSantos2020}, dipolar coupling\cite{chen_excitation_2019}, and an external magnetic field\cite{jamali_spin_2013}. Following the ideas behind graded-index optics, a continuous modulation of the magnetic parameters has been recently proposed  to control spin wave propagation\cite{davies_graded-index_2015,davies_towards_2015,hata_micromagnetic_2015,Tartakovskaya2020,Laurenson2020}. For example, it has been shown that a gradual modulation of the saturation magnetization ($M_s$) created with thermal landscapes can steer spin waves and change their dispersion relation as they propagate \cite{vogel_control_2018,vogel_optically_2015,borys_scattering_2019,mieszczak_anomalous_2020,gallardo_spin-wave_2019,kolokoltsev_hot_2012}. \\
	
	In this work, we use micromagnetic simulations to demonstrate that exchange spin waves do not propagate reciprocally in a perpendicularly magnetized ferromagnetic thin film in which Ms varies linearly along the length of the film.  To understand the origin of the phenomenon, we solve the linearized Landau-Lifshitz (LL) equation motion analytically.  The linear variation in $M_s$ along the $x$-direction creates an effective linear potential $V(x)$ in the spin wave equation of motion.  We transform to a non-inertial frame of reference where the inertial force cancels the force associated with the linear spin wave potential and allows the LL equation to be solved in the familiar constant $M_s$ condition.  However, when we transform back to the inertial frame, there is a frequency shift due to the acceleration of the excitation source similar to what happens in the Doppler effect that broadens the spin wave dispersion.  It is this Doppler shift of the spin waves that is the origin of the non-reciprocal behavior.

\begin{figure}
		\includegraphics[width=0.64\linewidth]{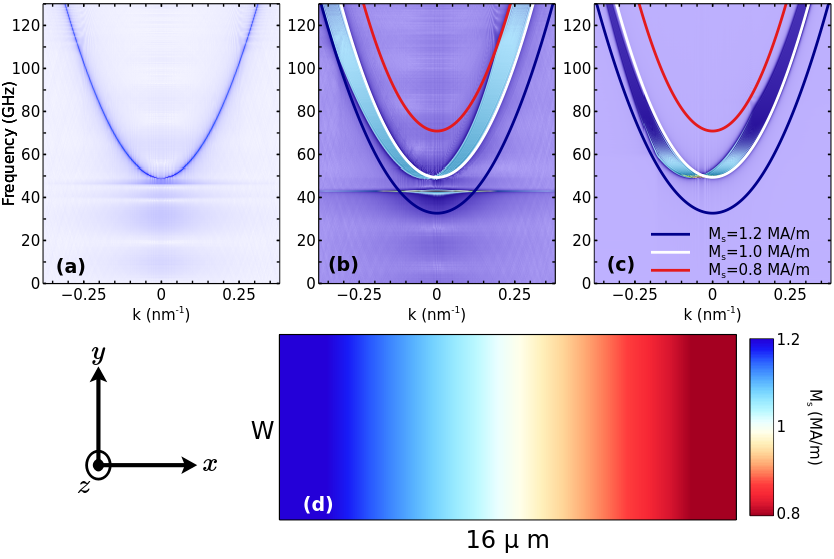}
		\caption{a) and b) are the dispersion curves found from micromagnetic simulations in the constant $M_s$ and gradient $M_s$ case, respectively. In b) solid lines represent the theoretical dispersions calculated using $M_s$ values at the edges and in the middle of the film. c) Dispersion curve obtained from the theoretical model with corresponding theoretical curves for comparison with b). e) $M_s$ linear gradient in the film used in the simulation. }
		\label{Fig:MsDisp}
	\end{figure}

	
	Using GPU-accelerated, micromagnetic code Mumax3\cite{vansteenkiste_design_2014}, we considered a $20$ $\mu$m $\times$ $256$ nm $\times$ $1$ nm film discretized using $10000$ $\times$ $128$ $\times$ $1$ finite difference cells. Periodic boundary conditions were used along the $y$ direction so that the effective width was $5376$ nm. We used magnetic parameters of perpendicular materials such as Pt/CoFeB\cite{zhou_temperature_2020}: exchange constant $A=15$ pJ/m, uniaxial anisotropy $K_u=1$ MJ/m$^3$ and applied field $\mu_0 H=1$ T, and recorded $m_y(x,t)$ in response to a field excitation of the form $h_0\sinc(2\pi f_c t)\hat{\mathbf{y}}$ with $\mu_0h_0=50$ mT and cutoff frequency $f_c=500$ GHz applied along the width over one cell in the $x$ direction positioned at the center of the film, $x=0$. The dispersion curve is obtained by performing a Fast Fourier Transform (2D-FFT) on $m_y(x,t)$ to get $m_y(k,\omega)$\cite{kumar_techniques_2017,venkat_proposal_2013}. We first considered a constant saturation magnetization $M_0=1$ MA/m throughout the sample and show the dispersion as a surface plot of $m_y(k,\omega)$ in Fig \ref{Fig:MsDisp} (a). The dispersion curve exhibits the typical exchange-driven quadratic form, $\omega_c\propto k^2$, in which spin waves propagating to the right and to the left have the same frequency. The magnetization gradient was modeled by a linear variation of $M_s$ across 250 regions in the range $x = [-8 \mu$m, $8 \mu$m] (See Fig \ref{Fig:MsDisp} (e)). The maximum value, $Ms(-8\mu m)=1.2$ MA/m, and minimum value, $Ms(8\mu m)=0.8$ MA/m may be achieved in Pt/CoFeB by creating a suitable thermal landscape\cite{zhou_temperature_2020}. In Fig \ref{Fig:MsDisp} (b) we show the dispersion curve obtained for spin waves propagating in a film with a $M_s$ gradient; additionally, solid lines indicate the theoretical dispersions corresponding to the $M_s$ values at the edges and the middle of the film. There is a horizontal line below the ferromagnetic resonance at $41$ GHz related to a strong spin wave localization at the samples edges due to the formation of a potential well in an inhomogeneous internal magnetic field\cite{jorzick_spin_2002}. Two features contrast the constant $M_s$ case: First, there is an asymmetry in the dispersion curve with respect to $k=0$. Second, the dispersion curve is significantly broadened. For positive (negative) propagation, $k>0$ ($k<0$), as the absolute value of $k$ increases, the broadening extend from the $1$ MA/m curve, white solid line, towards the $0.8$ MA/m red ($1.2$ MA/m blue) solid line.  \\

\begin{figure}
		\includegraphics[width=0.4\linewidth]{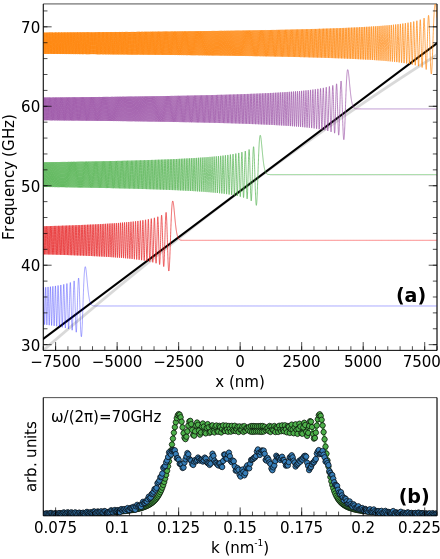}
		\caption{Stationary Solutions. In a) the solution for five different frequencies are presented, the black solid line corresponds to the potential considered for the theoretical calculations, while the light gray curve is the complete potential $\omega_0+qx+\gamma\mu_0(\alpha x)^2/M_0$. In b) we present the $k$ profile of the stationary solutions together with $m_y(k,t_0)$ from simulations for comparison.  } 
		\label{Fig:Stationary}
	\end{figure}

	To understand the dispersion curve, we construct an analytical model of the spin wave propagation in an uniaxial ferromagnetic film with magnetic energy, 
	
	\begin{align}
	E=\int\,dV\,A(\nabla \mathbf{m})^2-K_0\,m_z^2+\mu_0Hm_z,
	\end{align}
	
	where $K_0=K_u-\mu_0M_s^2/2$ is the effective anisotropy including the perpendicular demagnetizing field in the local approximation. We are interested in the dynamic behavior of the spin wave fluctuations, $\delta\mathbf{m}(\mathbf{x},t)$, around the static configuration, $\mathbf{m}_0=m_z$. To obtain the equations of motion in the long-wavelength limit we use $\mathbf{m}(\mathbf{x},t)=m_z+\delta\mathbf{m}(\mathbf{x},t)$ to linearize 
	
	\begin{align}\label{Eq:LL}
	\frac{\partial\mathbf{m}}{\partial t}=-\gamma\mathbf{m}\times\left(-\frac{1}{M_s}\frac{\delta E}{\delta\mathbf{m}}\right),
	\end{align}
	which can be cast as 
	\begin{align}\label{Eq:LL1}
	i\frac{\partial}{\partial t}m_+=\gamma\left[-\frac{2 A}{M_s}\frac{\partial^2}{\partial_x^2}+\frac{2 K_0}{M_s}+\mu_0H\right]m_+
	\end{align}
	for circularly polarized waves, $m_+=\delta m_x+i\delta m_y$. The saturation magnetization varies linearly as $M_s(x)=M_0+\alpha x$ which converts (Eq \ref{Eq:LL1}) to
	
	\begin{align}\label{Eq:LL2}
	i\frac{\partial}{\partial_t}m_+=\left[-\frac{1}{2\beta}\frac{\partial^2}{\partial_x^2}+\omega_0+q'x-\frac{\gamma\mu_0}{M_0}(\alpha x)^2-i\frac{\alpha x}{M_0}\frac{\partial}{\partial_t}\right]m_+
	\end{align}
	where we have defined the effective mass: $\beta=M_0/(4\gamma A)$, the effective potential $\omega_0=\gamma(2K_0/M_0+\mu_0H-\mu_0M_0)$, and $q'=\gamma\mu_0\alpha(H-2M_0)/M_0$ related to the force that the magnetization gradient exerts on the spin waves. Eq (\ref{Eq:LL2}) is a Schr\"odinger-like equation where the term quadratic in $x$ slightly modifies the linear potential and will be neglected (see Fig. \ref{Fig:Stationary} where the soft gray curve shows the effect of considering this term). The last terms couples the space and time coordinates. To continue with an analytical description, we replace the time derivative with the lowest possible spin wave frequency, the ferromagnetic resonant frequency, $\omega_0$. Then the equation to solve is

	\begin{align}\label{Eq:LL3}
	i\frac{\partial}{\partial_t}m_+=\left(-\frac{1}{2\beta}\frac{\partial^2}{\partial_x^2}+\omega_0+qx\right)m_+,
	\end{align}
	where $q=\gamma\mu_0\alpha/M_0(H-2M_0-\omega_0/\gamma\mu_0)=-\alpha\gamma\mu_0(1+2K_u/\mu_0M_0^2)$. It is worth  noting the importance of the space-time coupled term:, without it $q=\gamma\mu_0\alpha/M_0(H-2M_0)$, which would allow a change of sign for $H>2M_0$ and a fixed $\alpha$ value. The coupled term prevents the unphysical situation where the sign of $q$ is not determined entirely by $\alpha$.\\
	
	From the dispersion curve, Fig \ref{Fig:MsDisp} (b), it is clear that a function of the form $\omega(k)$ is not achievable in the presence of a magnetization gradient. To obtain an analytical description of the dispersion we perform a Fourier analysis of the solutions $m_+(x,t)$ to Eq. \ref{Eq:LL3}. We start with the Landau-Lifshitz equation that describes the spin waves in a perpendicularly magnetized magnetic film with a constant $M_s$ throughout the film,
	\begin{align}\label{Eq:Free}
	\left(-\frac{1}{2\beta}\frac{\partial^2}{\partial x'^2}+\omega_0\right)n_+=i\frac{\partial}{\partial t'} n_+
	\end{align}
	where $n_+(x',t')=n_x+in_y$, $n_x$ and $n_y$ are spin wave components, and the dispersion can be calculated to be $\omega_c(k)=1/(2\beta)k^2+\omega_0$. We then transform Eq \ref{Eq:Free} into an accelerated system described by $x=x'-1/2(q/\beta)t'^2$ and $t'=t$ with the acceleration of the system given by $-q/\beta$. Under this transformation, the derivatives are $\partial_x'=\partial_x$ and $\partial_t'=\partial_t-(q t/\beta)\partial_x$ so that the equation in the accelerated reference frame is
	\begin{align}\label{Eq:FreeTransf}
	-\frac{1}{2\beta}\frac{\partial^2}{\partial x^2}n_++\omega_0n_++i\frac{q t}{\beta}\frac{\partial}{\partial x}n_+=i\frac{\partial}{\partial t} n_+.
	\end{align}
	Eq. \ref{Eq:FreeTransf} rightly describes the spin waves in the transformed system. However, to an observer at rest in the accelerated reference frame, there should be a potential of the form $f x$ where $f$ is the inertial force producing the acceleration $-q/\beta$ instead of the coupled term $i (qt/\beta)\partial_xn_+$. Following refs. \cite{berry1979,greenberger_coherence_1979,greenberger_comment_1980,feng_complete_2001} we perform a unitary transformation 
	\begin{align}
	n_+(x',t')=m_+(x,t)e^{iqtx}e^{iq^2t^3/(6\beta)}
	\end{align}
	where $m_+(x,t)$ obeys Eq. \ref{Eq:LL3}	and effectively represents the physical situation with the potential $qx$ included. \\
	
	The stationary solution to Eq. \ref{Eq:LL3}, 
	\begin{align}
		Ai[(2\beta q)^{1/3}(x-x_0)]e^{-i\omega t}=Ai[B\xi]e^{-i\omega t},
	\end{align}
	is an Airy function with $x_0=(\omega-\omega_0)/q$, and is presented in Fig. \ref{Fig:Stationary} (a) for five different frequencies, $f=2\pi\omega$. In Fig. \ref{Fig:Stationary} (b) we present the $k$ profile for the stationary solution with $f=70$ GHz and compare with data obtained from a simulation in which the excitation field is of the form $h_0\sin(2\pi f_c t)\hat{\mathbf{y}}$ with $\mu_0h_0=50$ mT and frequency $f=70$ GHz. As a result of the $M_s$ gradient, one frequency excites a band of wavenumbers which in turn broadens the dispersion relation. \\
	
\begin{figure}
		\includegraphics[width=1\linewidth]{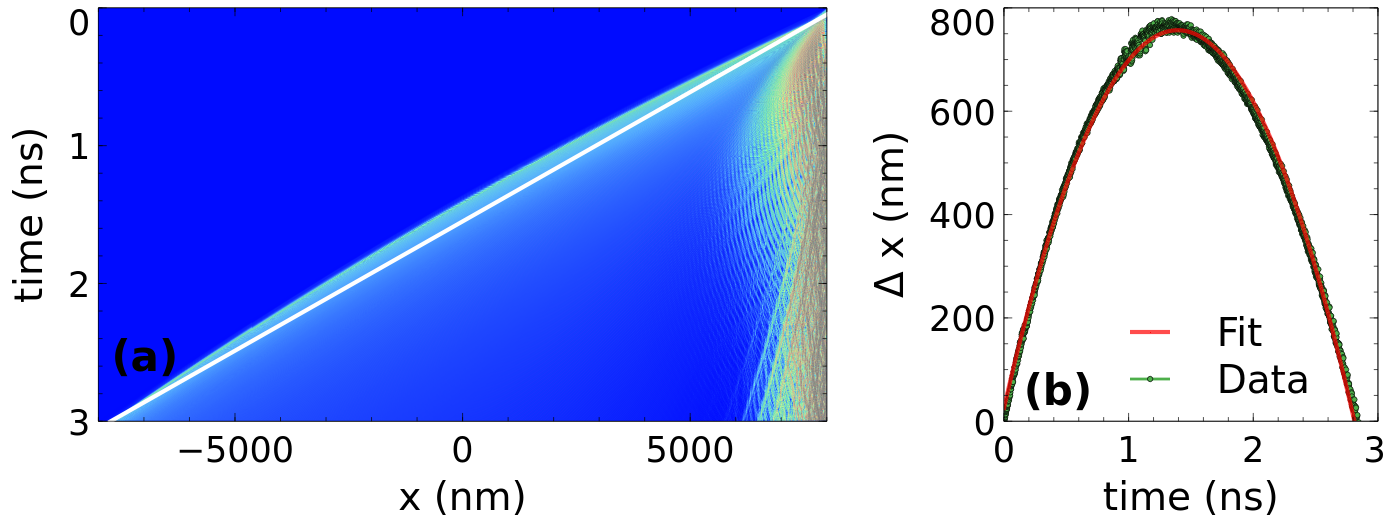}
		\caption{Spin wave acceleration. a) micromagnetic simulation of spin waves excited at the right edge of the $M_s$ gradient at $t=0$, the white solid lines represents the trajectory the front wave follows in the $M_s$ constant case. In b) we present the absolute difference $\Delta x $ between the front wave position in the $M_s$ gradient case and the front wave position in the $M_s$ constant case as a function of time. The red solid line corresponds to the fitting.  } 
		\label{Fig:acceleration}
	\end{figure}

	Using the stationary solution together with the integral representation of the Airy function, 
	\begin{align}
	Ai[B\xi]e^{-i\omega(k) t}=\frac{1}{2\pi B}\int_{-\infty}^{\infty}\;dk\;\exp[i(\frac{k^3}{3B^3}+k\xi-\omega(k)t)],
	\end{align}
	it is possible to construct an Airy wave packet. While we do not know $\omega(k)$ in the accelerated system, we can transform back to the primed, inertial, reference frame in which $\omega(k)=\omega_c(k)=1/(2\beta) k^2 +\omega_0$ to calculate the integral, 
	\begin{align}
	\frac{e^{-iqt(x'-qt'^2/2\beta)}e^{-iq^2t'^3/6\beta}e^{-i\omega_0t'}}{2\pi B}\int_{-\infty}^{\infty}dk\;\exp[i(\frac{k^3}{3B^3}+k\xi'-\frac{k^2 t'}{2\beta})],
	\end{align}
	using the useful formula $\int_{-\infty}^{\infty}du\,\exp[i(u^3/3+su^2+ru)]=2\pi e^{is(2s^2/3-r)}Ai(r-s^2)$. After transforming back to the accelerated system, the Airy wave packet becomes
	\begin{align}\label{Eq:AiryWP}
	m_+(x,t)=e^{-iqtx}e^{-iq^2t^3/6\beta}e^{-i\omega_0t}Ai\left[B(x+\frac{qt^2}{2\beta})-\left(\frac{B^2t}{2\beta}\right)^2\right]e^{-iB^2t/(2\beta)(B^4t^2/(6\beta^2)-B(x+\frac{qt^2}{2\beta}))}.
	\end{align}
	Substitution of Eq \ref{Eq:AiryWP} in Eq \ref{Eq:LL3} verifies it is a solution. Fig \ref{Fig:MsDisp} (b) shows the FFT of $m_+(x,t)$ obtained from Eq \ref{Eq:AiryWP} and displays a good agreement with the dispersion curve obtained from the micromagnetic simulation, Fig \ref{Fig:MsDisp} (b). In particular, the asymmetry and limits of the dispersion curve match. For higher frequencies our theoretical model appears narrower compared to the simulations. This is because of the approximation made on the space-time coupled term, Eq. \ref{Eq:LL2}. To visualize the accelerated reference frame and to compare the theoretical and simulated accelerations, we change the place of excitation from the middle to the right edge of the film and record $m_y(x,t)$ for the gradient and constant $M_s$ situations. Fig \ref{Fig:acceleration} (a) shows the recorded data for the $M_s$ gradient case and the solid white line corresponds to the position of the front wave in the constant $M_s$ case. The spin waves propagating in the $M_s$ gradient accelerate in the negative direction. The transformations are $x=x'-1/2(q/\beta)t'^2$, $t=t'$ where $(x,t)$ are the coordinates in the accelerated frame, and $(x',t')$ are the coordinates in the inertial system. The point $x=0$ corresponds to $x'=1/2(q/\beta)t'^2$ so that the accelerated frame is moving in the $1/2(q/\beta)t'^2$ direction. An observer in the accelerated frame, should feel an inertial force in the $-1/2(q/\beta)t'^2$ direction producing an acceleration $(-q/\beta)=-1.55\times10^{11}$ m/s$^2$ with the parameters used in the simulations. In Fig \ref{Fig:acceleration} (b) we show the difference between the front waves of spin propagating in the accelerated frame and in the inertial frame as a function of time. After fitting the curve we find that $\Delta x(t)=1/2a(t-t_0)^2-\Delta x_0$ with an acceleration $a=-7.52\times10^{11}$ m/s$^2$ and a time $t_0=1.40$ ns at which the maximum separation in the front waves,$\Delta x_0=0.76$ $\mu$m is reached. The theoretical acceleration, $q/\beta$, is lower than $a$ by a factor of five which is attributable to the two approximations being made, namely, the quadratic term in $x$ in the potential that was neglected, and the space-time coupled term that was replaced with the lowest possible frequency $\omega_0$. Still, the theoretical and simulated dispersion curves show a good agreement and the spin wave acceleration is clear.   	  \\

\begin{figure}
		\includegraphics[width=0.6\linewidth]{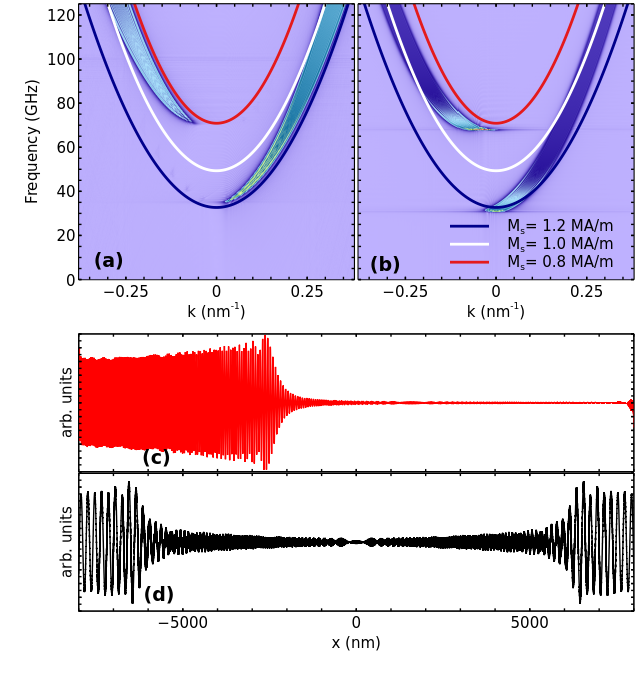}
		\caption{Unidirectional propagation of spin waves. a) Dispersion relation found from micromagnetic simulations where spin waves are excited at the edges of the $M_s$ gradient region. A discontinuity is found near $k=0$. b) Snapshot of the $m_y$ component of the magntization taken at $t=3$ ns after the start of the excitation where in the upper panel a $M_s$ gradient is considered and unidirectional propagation is achieved. For comparison the lower panel shows the constant $M_s$ case where reciprocal propagation is exhibited.  }
		\label{Fig:unidirectional}
	\end{figure}

	The situation changes when instead of exciting in the middle of the film, the excitation is made at the edges of the film. We recorded the $m_y(x,t)$ component in response to field excitations of the same form as above but now placed at the edges of the film, $x=-8\mu$m, and $x=8\mu$m, and the same for the remaining magnetic parameters. The dispersion curve is presented in Fig \ref{Fig:unidirectional} (a). As we only record the magnetic component, $m_y$ within the gradient region, spin waves excited at the left edge, $x=-8\mu$m, only propagate to the right. The $k>0$ branch of the dispersion is delimited by the $M_s=1.2$ MA/m and broadens to span the range of dispersions determined by the $M_s$ values within the film. Similarly,spin waves excited at the right edge, $x=8\mu$m, propagate to the left, with the $k<0$ branch bounded by the $M_s=0.8$ MA/m dispersion curve and their dispersion broadens towards the dispersion for $M_s=1.2$ MA/m. As a result, a discontinuity in the dispersion curve, shown in Fig. \ref{Fig:unidirectional} (a), is formed at $k=0$ and creates a frequency gap between right and left propagating states that can be calculated as the difference between the ferromagnetic resonances of the delimiting dispersion curves, 
	
	\begin{align}\label{Eq:deltaf}
		\Delta \omega=\mu_0\gamma(M_s^--M_s^+)\left(\frac{2K_u}{\mu_0M_s^+M_s^-}+1\right)
	\end{align}
	where $M_s^{\pm}$ corresponds to the delimiting $M_s$ value for the positive or negative dispersion branch. With our parameters we find $\Delta f=\Delta\omega/2\pi=37.5$ GHz. In our theoretical model the situation is described by including $x_0=\pm8$ $\mu$m in Eq. \ref{Eq:LL3} which modifies the Airy wave packet by a shift in the argument of the Airy function and a modification of the phase by a factor $e^{\pm iB^2t/(2\beta)Bx_0}$. Fig \ref{Fig:unidirectional} (b) shows the dispersion curve obtained from the theoretical model. The evident downward shift of the dispersion when compared to the simulation can be explained in terms of the neglected quadratic $x$ term in the potential that becomes larger at the edges of the gradient region. A key consequence of the discontinuity is that within the gap only one direction of propagation is permitted depending on the sign of the $M_s$ gradient. To verify, we again excite spin waves at the edges of the $M_s$ gradient region but change the form of the excitation to a sinusoidal field $\mathbf{h}(t)=h_0\sin(2\pi f t)\hat{\mathbf{y}}$ with $\mu_0h_0=50$ mT and a fixed frequency $f=50$ GHz which is in the middle of the frequency gap. In Fig. \ref{Fig:unidirectional} (b) we present a snapshot taken at $t=3$ ns: Propagation to the right is allowed while propagation to the left is forbidden. To compare, Fig. \ref{Fig:unidirectional} (c) shows what happens in the $M_s$ homogeneous case where propagation is reciprocal. \\
	
	Our results demonstrate that a $M_s$ gradient induces a nonreciprocal propagation of spin waves in a perpendicularly magnetized ferromagnetic film. The $M_s$ gradient is described by  an additional linear potential as compared to the constant $M_s$ case. Mathematically, the linear potential appears when transforming the constant $M_s$ case to an accelerated reference frame with acceleration $-q/\beta$. The asymmetry in the dispersion is then explained as a Doppler effect.  While non-reciprocity is observed in magnetostatic waves in thick films ($\approx$ $50$ $\mu$ m)\cite{kim_spin_2016,stancil2009spin}, the non reciprocity presented in this work can be achieved in films that are in the nano scale in thickness. Finally, we demonstrate that unidirectional spin wave propagation is achievable for a frequency band that depends on the $M_s$ gradient extreme values. Unidirectional propagation of exchange spin waves is of the highest importance for the design of magnonic computing devices. Our results are given in terms of a $M_s$ gradient that can be achievable through different methods, e.g. ion implantation\cite{marko2010determination,mcgrouther2005nanopatterning,fassbender2008magnetic}. However, the relevance of our study increases in light of recent studies in which modulation of the $M_s$ parameter is realized via a thermal landscape. We used parameters that correspond to the expected variation of the saturation magnetization in a temperature range of $0$-$300$ K in Pt/CoFeB. While the underlying physical mechanism is different, in practice, achieving unidirectional propagation by reversing the $M_s$ gradient resembles the working principle of a diode. Lastly, we have also verified that our results hold in the case where $M_s$ is constant throughout the film and the external magnetic field varies linearly.\\ 
	\textit{Note added}. During the final preparation of this manuscript we became aware of recently reported work on similar effects through spatially varying exchange (R. Macedo et al., MMM 2020 virtual conference, paper ER-04).

\begin{acknowledgments}
 This work was partially supported by fellowship Beca UNAM postdoctoral, Mitacs Globalink Research Award, National Council of Science and Technology of Mexico (CONACyT) under project 253754 and CB A1-S-22695, PAPIIT IG100519, Natural Sciences and Engineering Research Council of Canada (NSERC)
 
\end{acknowledgments}
	
\section{References}
\bibliography{Asymmetric.bib}
	
\end{document}